\documentclass[useAMS,usenatbib]{mn2e}

\usepackage{graphicx}

\def\be{\begin{equation}}
\def\ee{\end{equation}}
\def\ba{\begin{eqnarray}}
\def\ea{\end{eqnarray}}

\def\ltsima{$\; \buildrel < \over \sim \;$}
\def\simlt{\lower.5ex\hbox{\ltsima}}
\def\gtsima{$\; \buildrel > \over \sim \;$}
\def\simgt{\lower.5ex\hbox{\gtsima}}
\def\etal{{et al.\ }}

\usepackage[dvipsnames]{color}

\title[Nonequilibrium ionization states and cooling rates]
{Nonequilibrium ionization states and cooling rates of the photoionized enriched gas}
\author[E. O. Vasiliev]
       {Evgenii O. Vasiliev$^1$\thanks{E-mail:eugstar@mail.ru}\\
$^1$Institute of Physics, Southern Federal University, Stachki Ave. 194, Rostov-on-Don, 344090 Russia\\
}
\begin{document}
\date{Accepted 3004 December 15.
      Received 2004 December 14;
      in original form 2004 December 31}
\pagerange{\pageref{firstpage}--\pageref{lastpage}}
\pubyear{3004}
\maketitle

\label{firstpage}

\begin{abstract}
Nonequilibrium (time-dependent) cooling rates and ionization state calculations are 
presented for low-density gas enriched with heavy elements (metals) and photoionized 
by external ultraviolet/X-ray radiation. We consider a wide range of gas densities and 
metallicities and also two types of external radiation field: a power-law and the extragalactic 
background spectra. We have found that both cooling efficiencies and ionic composition
of enriched photoionized gas depend significantly on the gas metallicity and density, 
the flux amplitude, and the shape of ionizing radiation spectrum. 
The cooling rates and ionic composition of gas in nonequilibrium photoionization models 
differ strongly (by a factor of several) from those in photoequilibrium due to overionization of the ionic 
states in the nonequilibrium case. The difference is maximal at low values of the ionization
parameter and similar in magnitude to that between the equlibrium 
and nonequilibrium cooling rates in the collisionally controlled gas. 
In general, the nonequilibrium effects are notable at $T\simlt 10^6$~K. In 
this temperature range, the mismatch of the ionic states and their ratios between the 
photoequilibrium and the photo-nonequilibrium models reach a factor of several. The net result is
that the time-dependent energy losses due to each chemical element (i.e. the contributions 
to the total cooling rate) differ singificantly from the photoequilibrium ones.
We advocate the use of nonequilibrium cooling rates and ionic states for 
gas with near-solar (and above) metallicity exposed to an arbitrary ionizing radiation flux. 
We provide a parameter space (in terms of temperature, density, metallicity and ionizing radiation 
flux), where the nonequilibrium cooling rates are to be used. 
{ 
More quantitatively, the nonequilibrium collisional cooling rates and ionization states are 
a better choice for the ionization parameter log~$U \simlt -5$. The difference between the 
photoequlibrium and the photo-nonequilibrium decreases with the ionization parameter growth, 
and the photoequilibrium can be used for ionization parameter as high as log~$U\simgt -2$ 
for $Z\simlt 10^{-2}Z_\odot$ and log~$U\simgt 0$ for $Z\sim Z_\odot$. Thus, the nonequilibrium
calculations should be used for the ionization parameter range between the above-mentioned values.
In general, where the physical conditions are favour to collisional ionization, the nonequilibrium 
(photo)ionization calculations should be conducted. 
}
\end{abstract}



\section{Introduction}

\noindent

Recent observations of heavy elements (metals) in the intergalactic medium (IGM) give rich 
information about star formation and enrichment history of the Universe \citep{madau96,pettini,schaye03,nicastro,simcoe,danforth}.
In order to understand these observations, it is necessary to know which processes are 
responsible for the observed ionic composition and thermal state of the IGM. In the absence 
of ionizing background radiation, these processes are controlled by collisions and determined 
by gas metallicity and temperature. However, when galaxies and quasars form they produce
strong ionizing photons 
\citep{hm96,hm01,miniati,faucher}.
Therefore photoionization and photoheating are to be taken into account. 

Calculations of the cooling rates of astrophysical plasma in the collisional ionization
equilibrium (CIE) were performed by many authors 
\citep{house,cox,raymond,shull82,gaetz,bohringer,sd93,landi,benjamin,bryans}.
However, calculations of the time-dependent ionization of metals and associated radiative 
cooling showed significant deviations from the CIE states \citep{kafatos,shapiro,edgar,schmutzler,sd93,gs07}.
In some conditions, ionizing radiation can be important for the thermal 
and ionization states of primordial and enriched gas. \citet{efst} showed that the 
presence of UV radiation can strongly suppress the cooling rate of primordial gas. In the 
photoionization equilibrium case, \citet{wiersma} came to a similar conclusion for enriched 
gas exposed to the external ionizing radiation. It is obvious that ionizing radiation forces the 
ionic composition of gas to settle on to photoequilibrium. However, during transition from a pure 
collisional to a photoequilibrium case the nonequilibrium effects are expected to play a role. 

In general, numerical simulations of metal ionization states in the IGM should take into account the
effects of ionizing background. However, this requires tracing huge number of metal species 
and their ionization states, as well as complex calculations of the cooling and heating rates of the 
enriched photoionized gas. Therefore, simulations often consider either a  
limited number of metal ionization states or cooling rates of collisionally ionized gas. For 
instance, \citet{raga} studied a radiative bow shock using a limited ionization and 
cooling network; \citet{yoshikawa} used the CIE cooling rates to study the 
ionic composition of a warm-hot intergalactic medium. \citet{snneq} 
presented a comprehensive nonequilibrium ionization analysis of the supernova remnant.
Recently, self-consistent calculations 
were successfully performed in one-dimension. \citet{gs04} studied the metals
photoionization structures of pressure-supported gas clouds in dark matter minihalos and lately 
\citet{gs09} considered the ionic composition of gas in post-shock cooling layers 
behind fast radiative shock waves. 
More recently, three-dimensional nonequilibrium modelling of the interstellar meduim were performed
by \citet{avillez1,avillez2}. They considered the evolution of the turbulent magnetized 
interstellar meduim taking into account the time-dependent ionization of ten main chemical elements.
Here, we study how nonequilibrium affects the ionic composition and cooling rates of an enriched
photoionized gas and analyze a parameter space (i.e., temperature, density, metallicity, and ionizing 
radiation flux) where the nonequilibrium effects should be taken into account.

In the last decade studies of the intergalactic gas became very intensive. Among the most
interesting are the conclusions about the total mass of baryons in the Universe and its 
metal fraction. \citet{fukugita04} have estimated the cosmic baryon budget and found 
that less than 10\% of the total baryonic content in the Universe is locked in the form of collapsed objects 
(stars, galaxies, groups). More recently \citet{danforth} founded that approximately $\sim 50$\%
of the baryons remain unaccounted for. A significant discrepancy is also established for the total 
mass of metals observed in the Universe and the metal budget estimated from the total star formation rate. 
This lack of metals 
is known as the missing metals problem \citep{pettini}. Numerical simulations led by \citet{cen99} suggest 
that approximately 30 to 50\% of the cosmic baryons at $z=0$ are in the form of a diffuse intergalactic medium
with temperature of $10^5$~K$<T<10^7$~K, which has been called a warm-hot intergalactic medium (WHIGM).
Subsequent numerical simulations \citep[e.g.,][]{dave01} and more recent observational data
interpretation \citep{ferrara} also generally support this picture. 
However, a significant part of the baryons is still unaccounted for \citep{danforth}.
Thus, the question of missing baryons/metals physical condition is still open \citep{bregman}. 
Many current and future UV/X-ray missions
(HST\footnote{http://www.stsci.edu/hst/}, 
Chandra\footnote{http://chandra.harvard.edu/},
XMM-Newton\footnote{http://xmm.vilspa.esa.es/}, 
GALEX\footnote{http://www.galex.caltech.edu/}, 
FUSE\footnote{http://fuse.pha.jhu.edu/}, 
World Space Observatory -- WSO:  SPECT\-RUM-UV\footnote{http://wso.inasan.ru/},
SPECTRUM-X-GAMMA\footnote{http://hea.iki.rssi.ru/SXG/YAMAL/project\_eng.htm}, 
as well as XEUS\footnote{http://www.esa.int/science/xeus})
aim to find solution of it. In many numerical efforts for solving this 
question the equilibrium cooling rates \citep[e.g.,][]{cen06,yoshikawa,crain} are used. 
Obviously, feedback from galactic winds, supernova explosions and ionizing radiation from 
galaxies and quasars move the WHIGM away from the equilibrium. 
In this context nonequilibrium cooling 
rates and ionization states are a necessary component of the WHIGM ionic composition study.

The paper is organized as follows. In Section 2 we describe the details of the model and tests.
In Section 3 we present our results and discuss possible applications. In Section
4 we summarize our results.


\section{Atomic data and model description}

\noindent

We study the ionization and thermal evolution of a lagrangian element of 
cooling gas. A gas parcel is assumed to be optically thin for external ionizing 
radiation. In our calculations we consider all ionization states of the elements 
H, He, C, N, O, Ne, Mg, Si and Fe. We take into account the following major processes:
photoionization, multi-electron Auger ionization process, collisional ionization, 
radiative and dielectronic recombination
as well as charge transfer in collisions with hydrogen and helium atoms and ions. 
If we define the fraction of the ion stage $i$ as $X_i$, then the time-dependent 
evolution of $X_i$ is determined by
\ba 
 {dX_i \over dt} = - \gamma_i X_i - \alpha_i X_i n_e - \beta_i X_i n_e \\
   \nonumber
                    - (  \zeta_i^{\rm H} n_{\rm HI}  + \zeta_i^{\rm He} n_{\rm HeI}
                       +  \eta_i^{\rm H} n_{\rm HII} +  \eta_i^{\rm He} n_{\rm HeII}) X_i \\
   \nonumber
                                           + \alpha_{i+1} X_{i+1} n_e + \beta_{i-1} X_{i-1} n_e \\
   \nonumber
                    + ( \zeta_{i-1}^{\rm H} n_{\rm HI}  + \zeta_{i-1}^{\rm He} n_{\rm HeI})  X_{i-1} \\
   \nonumber
                    + (  \eta_{i+1}^{\rm H} n_{\rm HII} +  \eta_{i+1}^{\rm He} n_{\rm HeII}) X_{i+1} \\
   \nonumber
                    + \Sigma_{j<i} \gamma_{j\rightarrow i} x_j,
\ea
here $\gamma$ is the photoionization rate of each ion, given by
\be
 \gamma_i = 4\pi \int_{\nu_{i,0}}^\infty {J_\nu \over h\nu} \sigma_{i,\nu} d\nu,
\ee
and $\nu_{i,0}$ is the ionization frequency, $J_\nu$ is the background radiation flux,
$n_{\rm HI}, n_{\rm HII}, n_{\rm HeI}, n_{\rm HeII}$ and $n_e$ are the number densities
for neutral hydrogen, ionized hydrogen, neutral helium, ionized helium, and electrons, 
respectively, $\sigma_{i,\nu}$ is the photoionization cross section adopted from \citet{verner96}, 
\citet{vya95}, $\alpha$ is the recombination rate, including 
radiative recombination \citep{shull82,aray,verner96,badnell06a}, and dielectronic recombination 
\citep{badnell06b,bautista,colgan03,colgan04,altun04,altun06,z2003,z2004a,z2004b,z2006,mitnik,mazzotta,badnell06c}, 
and $\beta$ is the collisional ionization rate adopted from \citet{voronov}, 
$\eta^{\rm H}, \eta^{\rm He}, \zeta^{\rm H}, \zeta^{\rm He}$ are the charge transfer rates
of ionization and recombination with hydrogen and helium adopted from \citet{arot} and \citet{kingdon}, 
$\gamma_{j\rightarrow i}$ is the total photoionization rate including that from deep shells followed by 
Auger-electron ejection, the Auger effect probabilities are taken from \citet{kaastra}.

The system (1) must be complemented by the temperature equation.
Neglecting the change of number of particles in the system (for fully ionized hydrogen
and helium it remains approximately constant) the gas temperature is determined by 
\be
 {dT \over dt} = {n \Gamma - n_e n_H  \Lambda \over A n k_B}
\ee
where $n$, $n_e$ and $n_H$ are the electron and total hydrogen number 
densities, $\Gamma(x_i,T,Z,J_\nu)$, $\Lambda(x_i,T,Z)$ are heating and cooling 
rates, $A$ is a constant equal to $3/2$ for isochoric and $5/2$ for isobaric 
cooling, $k_B$ is the Boltzman constant. Cooling is isobaric when the 
cooling time is much greater then the dynamical time of a system, $t_c/t_d \gg 1$ 
and it is isochoric when the time ratio is opposite, $t_c/t_d \ll 1$. In the 
lack of a dynamic determination we are able to constrain the dynamical time only 
by the age of the universe, $t_H \sim 2/3H(t)$. Hot gas ($T>10^6$~K) with low 
overdensities, $\delta = (\rho - \langle\rho_b(z)\rangle)/ \langle\rho_b(z)\rangle\sim 1$,
has cooling time higher than the local Hubble time. In this case
the adiabatic approximation is a good choice. 

The total cooling and heating rates are calculated using the photoionization code CLOUDY 
\citep[ver. 08.00,][]{cloudy}. More specifically, we input into CLOUDY code a given 
set of all ionic fractions $X_i$ calculated at temperature $T$, gas number density $n$ and 
external ionization flux $J(\nu)$ and obtain the corresponding cooling and heating rates. 
For the solar metallicity we adopt the abundances reported by \citet{asplund}, except 
Ne for which the enhanced abundance is adopted \citep{drake}. The solar abundances 
are listed in Table~1. In all our calculations we assume the helium mass fraction 
$Y_{\rm He} = 0.24$.

We are interested in the ionization and thermal evolution of gas exposed to external ionizing 
radiation. A shape of ionizing background spectrum gives information about population of 
sources which give major contribution to the background. An exposed gas can be either located
both near galaxy or quasar, or far from any source. In the former the ionizing radiation is 
inherent to a particular source, in the latter the gas is radiated by the extragalactic background 
formed by the galactic and quasar population. 

\begin{figure}
\includegraphics[width=80mm]{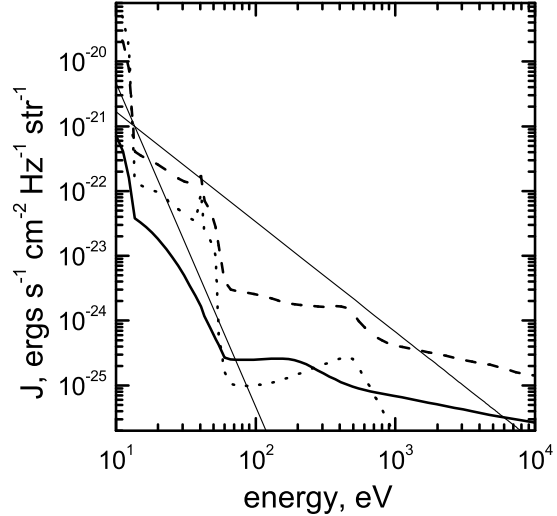}
\caption{
The ionizing backgrounds. The Haardt \& Madau (2001) spectra at redshifts
$z=0,\ 3,\ 6$ are depicted by thick solid, dashed and dotted lines correspondingly.
The power-law profiles with $J_{21}=1$ and the spectral index $\alpha$ = 1.7, 5 
are presented by thin solid lines.
}
\label{figspec}
\end{figure}

\begin{figure}
\includegraphics[width=80mm]{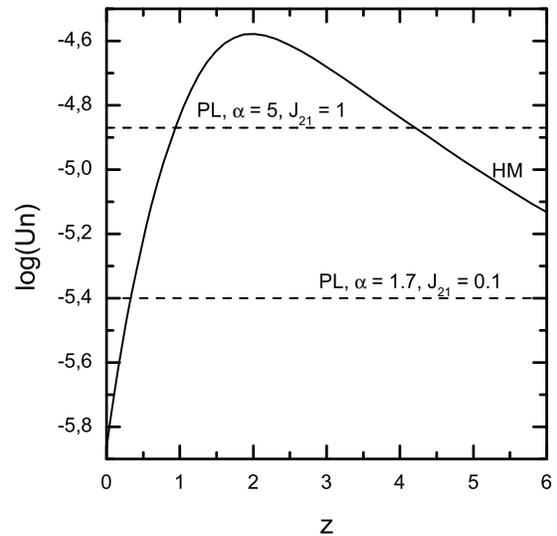}
\caption{
The parameter $Un$ ($U$ is the ionization parameter) for the Haardt \& Madau (2001) 
spectra (HM) -- solid line, and the power-law spectra (PL) -- dash lines.
}
\label{figu}
\end{figure}

Here we consider two types of spectra: a power-law spectrum, which is typical to stellar populations 
of galaxies, active galactic nuclei and quasars, and the extragalactic background calculated by using 
the radiative transfer in the cosmological simulations \citep[ HM01]{hm96,hm01}. A power-law (PL)
spectrum shape is $J = J_{21} (\nu/\nu_H)^{-\alpha}$, 
where $J_{21}$ is the flux value at the hydrogen ionization limit in units 
$10^{-21}$~erg~s$^{-1}$~cm$^{-2}$~str$^{-1}$, $\alpha$ is the spectral index. 
To study the photoionization of gas around a quasar $\alpha$ is taken equal to 1.7,
\citep{sazonov}, but we note that other studies have found both softer and harder 
spectra, with a significant variance about this value \citep{zheng,telfer,scott}.
We note that the quasar spectra show a gradual flattening in the 0.2--2~keV band \citep[e.g. ][]{sazonov}.
Such a property can be clearly seen for the cumulatuve extragalactic background for $E\simgt 1$~keV at
redshifts 0 and 3, where the contribution from quasars is dominant. We have also made calculations for shallower spectrum, $\alpha = 1$, because some authors use this value \citep[see e.g.][]{kitayama}.
To mimic the stellar cluster ionization background, we adopt the spectrum with $\alpha = 5$. For such a spectrum 
shape, the photon-number weighted average intensity above 13.6~eV is close to that for the blackbody spectrum
with an effective temperature $\sim 2\times 10^4$~K \citep{kitayama}. 
Certainly, to model an adequate stellar cluster spectrum the theoretical starburst calculations are necessary. 
However, such models depend on many parameters, e.g., initial mass function, star 
formation efficiency, etc. Therefore, we have adopted a simple power-law shape.
To model the influence from the extragalactic ionizing background 
we use the \citet{hm01} ionizing radiation flux, which is included in CLOUDY code. 
In our simulations we take the radiation spectrum from 1 to $10^4$ eV. Figure~\ref{figspec} presents the
ionizing background profiles used in our calculations. 
It is worth noting that the original spectra presented by HM01 do not include the absorption 
in the He II resonant lines \citep{madau09,faucher},
although such an absorption can change the abundances of metal ions like CIII and 
SiIV \citep{madau09} at redshifts before the helium reionization. 

To characterize an ionizing background it is convinient to use the ionization parameter $U$
defined as the dimensionless ratio of hydrogen ionizing photons to the total hydrogen density.
For a power-law spectrum with the spectral index $\alpha$, the ionization parameter can be easy 
found: $U = 6.8\times 10^{-5} J_{21} / n\alpha$. Figure~\ref{figu} shows the parameter $Un$,
where $n$ is the total number density, for the extragalactic ionizing background (HM01).

\begin{table}
\caption{Solar elemental abundances}
\center
\begin{tabular}{lc}
\hline
\hline
   Element  &    (X/H)$_\odot$  \\
\hline
Carbon      &  $2.45\times 10^{-4}$    \\
Nitrogen    &  $6.03\times 10^{-5}$    \\
Oxygen      &  $4.57\times 10^{-4}$    \\
Neon        &  $1.95\times 10^{-4}$    \\
Magnesium   &  $3.39\times 10^{-5}$  \\
Silicon     &  $3.24\times 10^{-5}$   \\
Iron        &  $2.82\times 10^{-5}$    \\
\hline
\hline
\end{tabular}%
\label{table1}
\end{table}

\begin{figure}
\includegraphics[width=80mm]{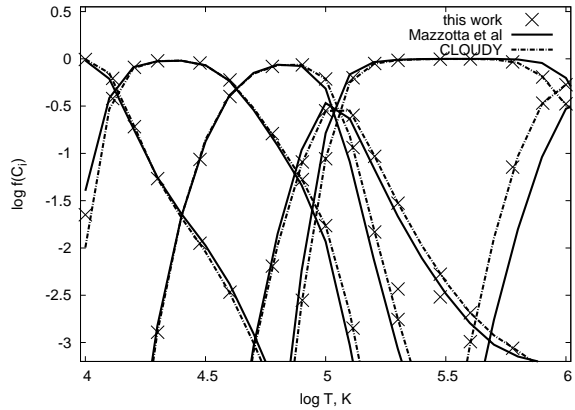}
\caption{
         The collisional equilibrium carbon state fractions 
         for our code (crosses) are superposed on the
         results by \citet{mazzotta} (solid lines) and the 
         standard collisional equilibrium test of CLOUDY (dash-dotted lines).
         }
\label{figcieC}
\end{figure}

\begin{figure}
\includegraphics[width=11cm]{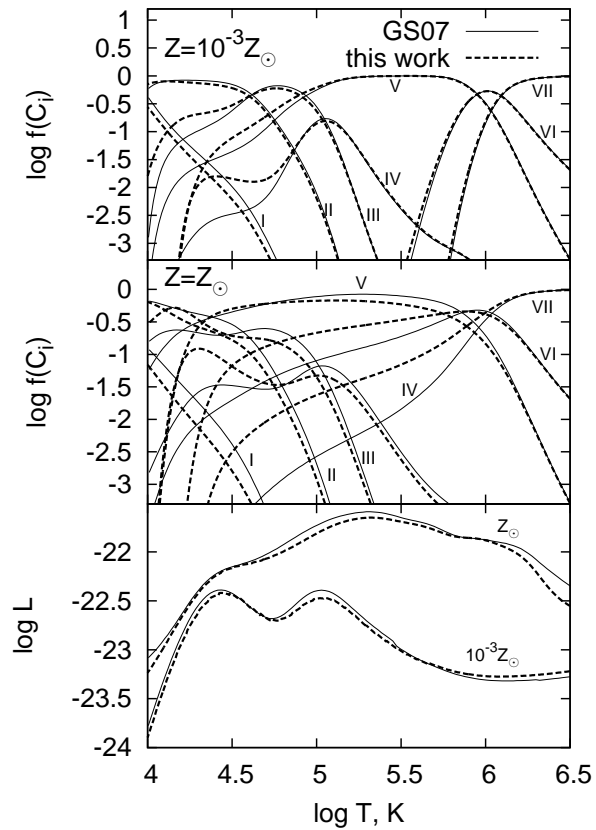}
\caption{
         The collisional time-dependent carbon state fractions 
         for our code (thick dashed lines) are superposed on the
         results by \citet{gs07} shown by thin solid lines
         for metallicities $10^{-3}Z_\odot$ (upper panel) and solar value (middle panel).
         The roman numbers indicate the ionizaion state.
         The collisional time-dependent cooling rates obtained from 
         our code (thick dashed lines) and \citet{gs07} data (thin solid lines)
         for $10^{-3}Z_\odot$ (lower curves) and solar (upper curves) metallicities
         are shown in the lower panel.
         }
\label{figneqC}
\end{figure}

We solve a set of 96 coupled equations (95 for ionization states and one for temperature)
using a Variable-coefficient Ordinary Differential Equation solver \citep{dvode}. 
The time step is chosen equal to the minimum value between ($0.1 t_{ion}, 0.01 t_{cool}$),
where $t_{ion}$ and $t_{cool}$ are the ionization and cooling times, correspondingly.

The evolution of gas certainly depends on a choice of the initial condition of gas, namely, 
the initial ionic composition and temperature. We consider such a dependence elsewhere 
\citep[some analysis can be found in][]{v10}. In this paper we start our calculations from 
temperature $T = 10^8$~K with the collisional equilibrium ion abundances. 
At the beginning we present several tests of the method of calculation. In these tests we also start 
our calculations from temperature $T = 10^8$~K with the collisional equilibrium ion abundances. 
During the calculation the gas number density is assumed to remain unchanged due to any dynamical 
or chemical processes. We stop our calculations when the temperature reaches $10^4$~K or 
photoionization heating rate differs from cooling rate by less than 5\% ($|\Gamma-\Lambda| 
/\Lambda < 0.05$). The ionic composition is almost "frozen" at this moment. Note that with such 
a stopping criterion a gas still remains "time-dependent" at the end of the calculation, as it 
has not yet reached ionization equilibrium.  

Figures~\ref{figcieC}-\ref{figneqC} present the main tests of our code used for the calculations. 
At first we tested our model by comparing calculations in the collisional equilibrium 
(we run the simulations long enough at constant temperature so that eventually all ion species 
settle in ionization equilibrium). 
Figure~\ref{figcieC} shows the carbon ionization fractions in the collisional equilibrium 
obtained from our code (depicted by crosses) and the results by \citet{mazzotta}
as well as the standard CIE test of the CLOUDY code. One can see a good coincidence between 
the results of our code and those obtained in the previous works. 

Two upper panels of Figure~\ref{figneqC} presents the carbon ionization fractions in the collisional
time-dependent case obtained from our code and the results by \citet{gs07}. One can 
see that the fractions in both calculations are close to each other for low metallicity value,
$Z = 10^{-3}Z_\odot$, while the strong deviations can be found for solar metallicity. We expect that 
this difference is due to various atomic data and, as a consequence, ionization and thermal 
evolution. Actually, \citet{gs07} have utilized the CLOUDY code ver.06.00, whereas 
in our calculations more recent version (08.00) of the code is used. There were several revisions 
of atomic data between these versions\footnote{http://wiki.nublado.org/wiki/RevisionHistory}. 
Also we don't include in our code the cooling rate by sulfur, but small variation in the total
cooling rate can produce different ionic composition. Thus, in low metallicity gas the ionic 
composition obtained in our code is close to that of \citet{gs07}, because
the thermal evolution is determined by hydrogen and helium. Whereas the increase of metallicity
leads to stronger deviation between our and \citet{gs07} results.
Note that the total cooling rates in both calculations are close, the rates are shown in
the lower panel of Figure~\ref{figneqC}.


\section{Results}

In this section we consider the isochoric cooling rates and the ionization states.
We leave out of scope of this paper the isobaric cooling, because the increase of gas density 
during isobaric process, obviously, leads to the collisionally controlled ionization states and 
radiative cooling (the increase factor of density is $10^4$ if we study the gas evolution
from $10^8$ to $10^4$~K). 

\begin{table*}
\caption{List of models}
\center
\begin{tabular}{ccc}
\hline
\hline
n, cm$^{-3}$&    $Z/Z_\odot$              &      spectrum      \\   
\hline
$10^{-4}$, $10^{-3}$, $10^{-2}$& $10^{-3} - 1$ & power-law (PL), $\alpha$ = 1, 1.7, 5; J = (10$^{-4}$--1)$J_{21}$\\
$10^{-4}$, $10^{-3}$, $10^{-2}$& $10^{-3} - 1$ & Haardt \& Madau (2001) data (HM), $z= 0 - 6$ \\
\hline
\hline
\end{tabular}%
\label{table1}
\end{table*}

\begin{figure*}
\includegraphics[width=12cm]{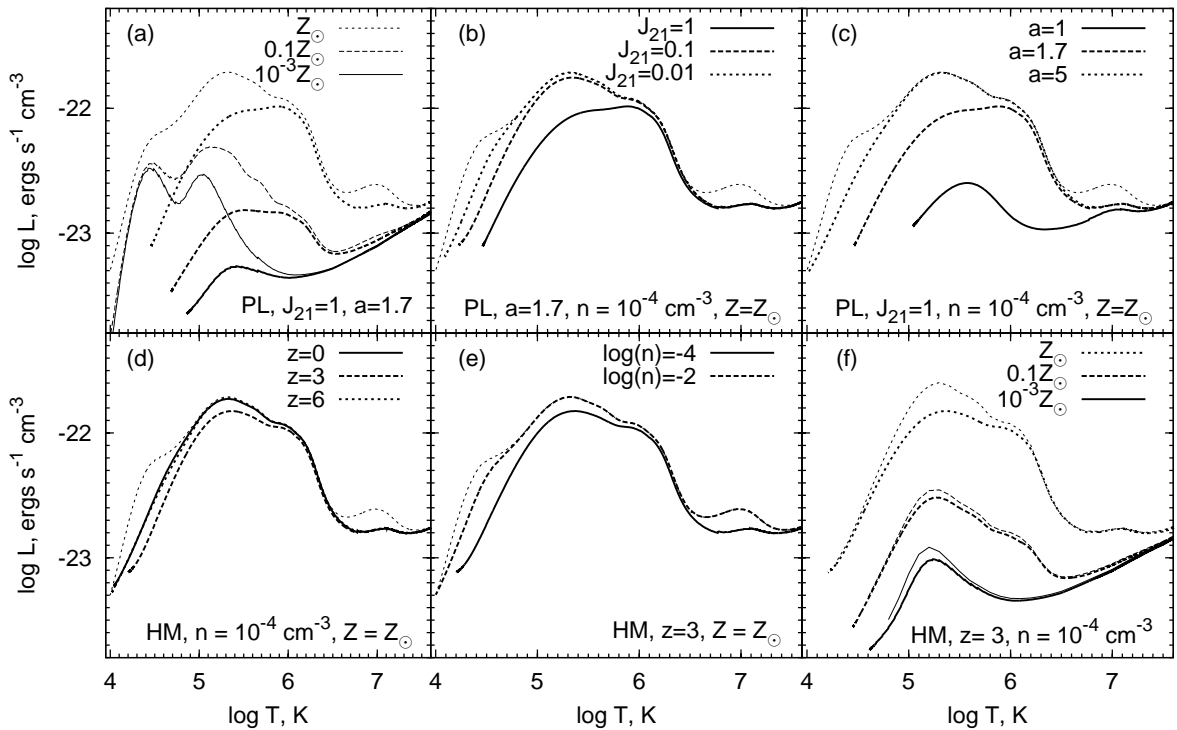}
\caption{
The cooling rates for gas exposed to ionizing radiation:
(a) the dependence on metallicity for gas density $n=10^{-4}$~cm$^{-3}$ of gas
exposed to the power-law (PL) external radiation with $J_{21}=1$ and $\alpha = 1.7$, 
the cooling rates for metallicity $Z = 10^{-3},\ 10^{-1}$ and 1 $Z_\odot$ 
are shown by thick solid, dashed and dotted lines, correspondingly,
the cooling rates for pure collisional case are depicted by thin lines of the same type,
(b) the dependence on flux amplitude $J_{21}$ for power-law spectrum with $\alpha=1.7$ 
and gas with $n=10^{-4}$~cm$^{-3}$, $Z = Z_\odot$, the cooling rates for 
$J_{21}=0.01,\ 0.1,\ 1$ are shown by thick dotted, dashed and solid lines, correspondingly;
(c) the dependence on spectral index $\alpha$ for gas with $n=10^{-4}$~cm$^{-3}$, $Z = Z_\odot$
exposed to $J_{21}=1$, the cooling rate for the spectral index $\alpha =1$ 
is shown by thick solid line, the rate for $\alpha=1.7$ -- thick dashed line, and 
the rate for $\alpha=5$ -- thick dotted line;
(d) the rates for gas with $n=10^{-4}$~cm$^{-3}$ and solar metallicity 
exposed to the \citet{hm01} background radiation (HM), the solid, dashed and 
dotted lines correspond to the backgrounds at redshifts $z=0, 3, 6$;
(e) the rates for gas with solar metallicity and number densities 
$n=10^{-4},\ 10^{-2}$~cm$^{-3}$ (thick solid and dashed lines, correspondingly)
exposed to the \citet{hm01} background radiation at $z=3$;
(f) the nonequilibrium (thick) and photoequilibrium (thin lines) 
cooling rates for gas with $n=10^{-4}$~cm$^{-3}$ and metallicities $Z = 10^{-3},\ 10^{-1},\ 1\ Z_\odot$ 
(solid, dash and dot lines, correspondingly) exposed to the \citet{hm01} background 
radiation at $z=3$, the photoequilibrium rates are similar to those obtained by \citet{wiersma}, 
see details in the text;
In all panels, except(a) and (f), the collisionally induced cooling rate for gas with 
$n=10^4$~cm$^{-3}$, $Z=Z_\odot$ is depicted by thin dotted line.
}
\label{figcool}
\end{figure*}

\subsection{Cooling rates}

In Table 2 we list the main parameters of the models considered here\footnote{  The  
cooling rates and ionic composition for several models
are available at  http://ism.rsu.ru/cool/cool.html.}. 
We start from temperature $T_i = 10^8$~K with the collisional 
equilibrium ion abundances. This seems to be a good approximation, because the ionic 
composition at so high temperature is mainly determined by frequent collisions.

\subsubsection{Dependence on metallicity for power-law spectrum}

Figure~\ref{figcool} shows the cooling rates for gas exposed to external 
radiation for several models listed in Table 2. In the panel (a) the dependence 
of cooling efficiency on the gas metallicity is presented. The other parameters are 
fixed: $n=10^{-4}$~cm$^{-3}$, $J_{21}=1$ and $\alpha=1.7$. In addition, the collisional 
nonequilibrium cooling rates for the same metallicities are depicted by thin 
lines. A substantial difference between the cooling rates of photoionized and  
collisional gas at temperature lower $10^6$~K is clearly seen. The cooling of 
photoionized gas is suppressed in contrast to that of collisionally ionized 
gas, because the gas is overionized due to photoionization. Indeed, one can see that 
the cooling rates of photoionized gas at $T\sim 10^5$~K are smaller by an order 
of magnitude in comparison with that of collisionally ionized gas. An increase of metallicity provides 
more effective cooling of gas and shifts the maximum in the cooling rate to higher temperatures. 
In general, for the same physical state of gas the photoionization cooling rates are lower 
than the collisional ones. 

\subsubsection{Dependence on flux amplitude and spectral index}

In the panel (b) one can see the cooling rates for gas exposed to radiation with fluxes 
$J_{21}=0.01,~0.1,~1$, the other parameters, $\alpha=1.7$, $n=10^{-4}$~cm$^{-3}$, 
$Z=Z_\odot$, were fixed. The cooling rate is getting suppressed significantly with increase of 
the flux. The panel (c) presents the cooling rates for three different values of
the spectral index $\alpha = 5,\ 1.7$ and 1. In this case the picture is similar to the 
previous one. For less steep spectrum the cooling rate is suppressed stronger.

\subsubsection{Dependence on redshift for the extragalactic spectrum}

The spectrum formed by a cumulative extragalactic radiation field from galaxies and quasars 
differs significantly from a simple power-law spectrum (see Figure~\ref{figspec}). Consequently, 
the cooling rate of gas exposed to the extragalactic radiation also differs from that 
exposed to a power-law spectrum. Figure~\ref{figcool}d presents cooling rates for gas with 
$n=10^{-4}$~cm$^{-3}$ and solar metallicity exposed to the HM01 extragalactic background 
spectra at redshifts $z=0, 3, 6$. Among these redshifts the ionizing radiation flux is the strongest 
at $z=3$, so a difference from the collisional cooling rate is the most significant at $z=3$. 
The cooling rates for two other redshifts presented here coincide between each other. In comparison 
with the collisional case the rates demonstrate notable differences at $T< 10^6$~K for $z=3$ and 
$T \simlt 8\times 10^4$~K for $z=0$ and 6.
Such differences are caused by the overionization of the dominant coolants. A more detailed 
discussion can be found in the next subsection.

\subsubsection{Dependence on density and metallicity for the extragalactic spectrum at $z=3$}

\begin{figure*}
\includegraphics[width=18cm]{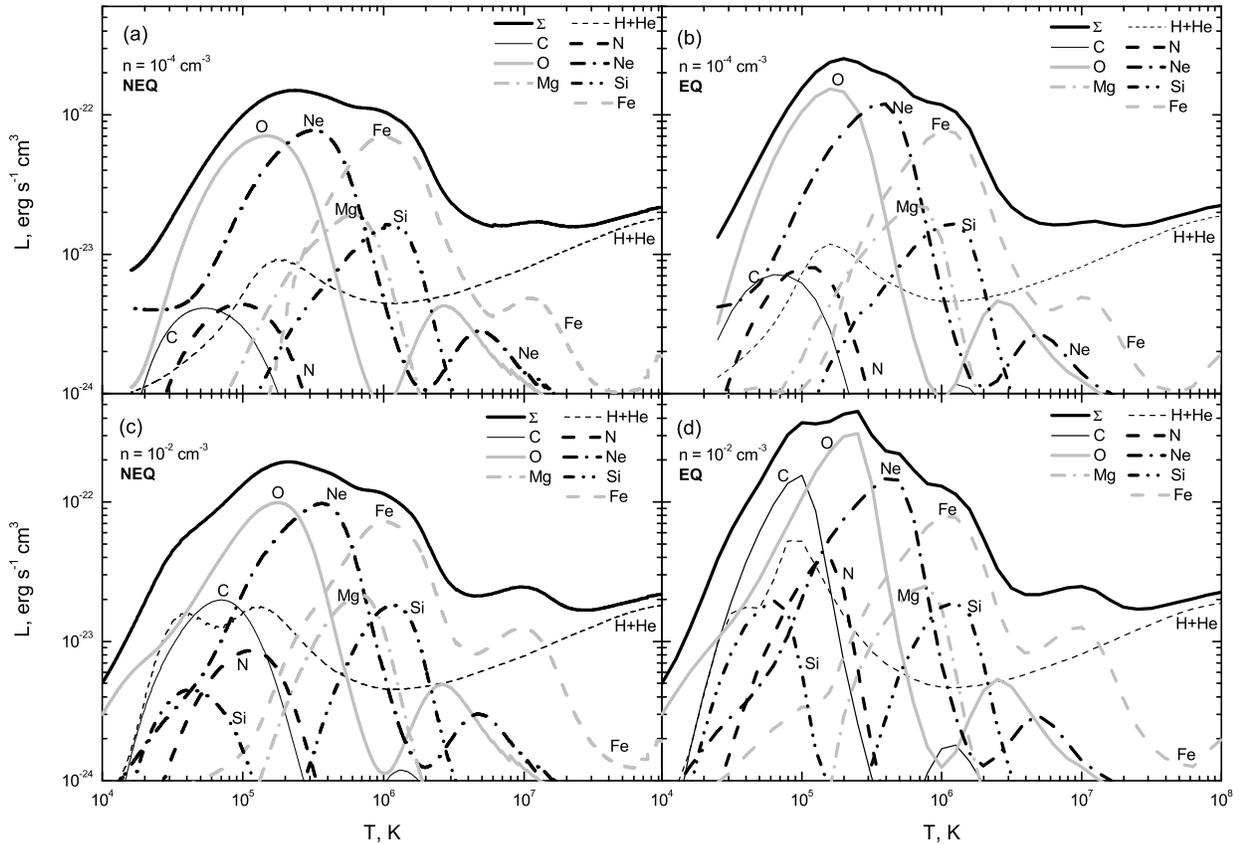}
\caption{
The cooling rates in photo-nonequilibrium (left panels) and photo-equilibrium (right panels) for
gas with solar metallicity and $n=10^{-4}$~cm$^{-3}$ (upper panels) and $n=10^{-2}$~cm$^{-3}$ (lower panels) , 
exposed to the extragalactic HM ionizing radiation at redshift $z=3$. 
The thick solid black line shows the total cooling rate, the thin dash line shows the contribution from 
hydrogen and helium, the other lines (see the legend on the panels) show the contributions from 
each chemical element.
}
\label{figcoolants}
\end{figure*}

Fugure~\ref{figcool}e shows the dependence of the cooling rates on gas density. This 
dependence is determined by the collisional ionization and photoionization time ratio,
which is proportional to $n_e/J_{21}$. A higher ionizing flux leads to a predominance 
of the photoionization, whereas an increase of density intensifies the collisional ionization.
{ 
For instance, the cooling rates of collisional and photoionized gas with $n=10^{-2}$~cm$^{-3}$ 
(panel e) are equal to each other at $T\simgt 5\times 10^4$~K. A small difference at lower temperature 
is determined by the overionization of hydrogen, carbon and oxygen, which are the main contributors
to the cooling rate in this temperature range.
}

Figure~\ref{figcool}f shows the dependence on metallicity of nonequilibrium cooling rates 
for gas with number density $n=10^{-4}$~cm$^{-3}$ exposed to the HM01 spectrum at $z = 3$. 
They differ considerably from the nonequilibrium collisional rates presented on the panel (a).
The presence of UV radiation leads to overionized ionic composition, consequently, suppresses 
the cooling rates. 

Recently \citet{wiersma} have calculated the cooling rates in the photoionization equilibrium. 
Because of the restricted number of the chemical elements and its different abundances in our model we 
recalculate photoequilibrium rates similar to \citet{wiersma} and compare it to 
the nonequilibrium photoionization cooling rates. We add the corresponding 
cooling rates in Figure~\ref{figcool}f. One can see a good coincidence between the equilibrium 
and the nonequilibrium cooling rates of photoionized gas for $Z\leq 0.1Z_\odot$. This 
coincidence is a result of the influence of the ionizing radiation, whereas for the collisional 
case the equilibrium and the nonequilibrium cooling rates differ significantly \citep[e.g.,][]{sd93}.
A sufficiently high ionizing radiation flux forces the ionic composition to settle 
onto the equilibrium, though a small difference is seen for solar metallicity at $T\sim 2\times 10^5$~K.
Therefore, one can expect that nonequilibrium effects will become more significant for higher metallicity.

\begin{figure*}
\includegraphics[width=19cm]{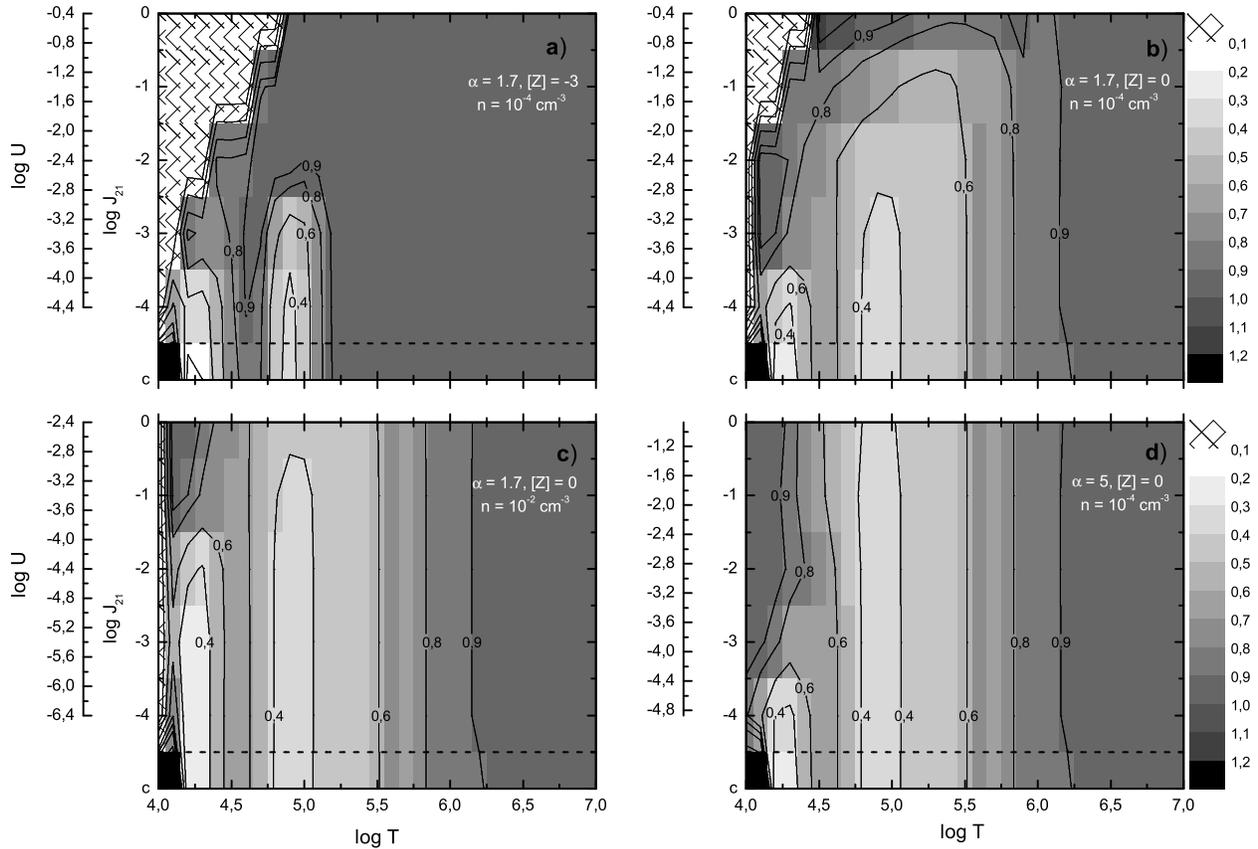}
\caption{
The ratio of nonequilibrium to equilibrium cooling rates for gas exposed to a power-law spectrum:
a) $n=10^{-4}$~cm$^{-3}$, $Z=10^{-3} Z_\odot$ and the index $\alpha=1.7$; 
b) $n=10^{-4}$~cm$^{-3}$, $Z=Z_\odot$ and $\alpha=1.7$; 
c) $n=10^{-2}$~cm$^{-3}$, $Z=Z_\odot$ and $\alpha=1.7$;
d) $n=10^{-4}$~cm$^{-3}$, $Z=Z_\odot$ and $\alpha=5$.
The ionization parameter $U$ corresponded to the spectrum and number density is shown by extra y-axis.
The letter "c" on the y-axis (or the area is separated by dash line) corresponds to the ratio 
of the cooling rates of collisional plasma (no photoionization). 
The shaded area corresponds to the conditions under which the photoheating exceeds the 
cooling in the nonequilibrium case.
}
\label{figrelpl}
\end{figure*}

Let us consider the difference between the nonequilibrium (time-dependent) and photoequilibrium cooling rates.
Figure~\ref{figcoolants} shows the individual contributions from each chemical element to the total
cooling rate for gas exposed to the HM01 extragalactic ionizing radiation at $z=3$ in the
photo-nonequlibrium (left panels) and photoequilibrium (right panels). Figure~\ref{figcoolants} presents
the rates for solar metallicty and two number densities, $10^{-4}$ and $10^{-2}$~cm$^{-3}$. 
The time-dependent rates coincide with the photoequilibrium ones at $T\simgt 10^6$~K. At the same time, for 
$T\simlt 10^6$~K the total cooling rates for both densities differ significantly in the 
photoequilibrium and time-dependent models. This difference comes from individual contributions 
of the considered elements. One can see that the contributions from C, N, O and Ne in the time-dependent case are 
smaller than those in the equilibrium. These elements are overionized in the time-dependent case. 
Such a picture is determined by the recombination timescales of the C, N, O and Ne ionic states, which
lead to the recombination lag in the time-dependent model. For $n=10^{-2}$~cm$^{-3}$ the {  hydrogen} 
and helium should be added to the list of the overionized ions. Panels (c-d) show a strong difference in the 
H\&He conributions. 
{ 
For $n=10^{-2}$~cm$^{-3}$ collisions dominate over photoionization in a wide temperature range, so the
photo-nonequilibrium cooling rate coincides with the collisional one at $T\simgt 5\times 10^4$~K
(see Figure~\ref{figcool}e) and the photoequilibrium and CIE cooling rates equal to each other
at $T\simgt 4\times 10^4$~K. Below $T\sim 4\times 10^4$~K the difference between photoequilibrium and CIE 
cooling rates is mainly determined by the overionization of hydrogen and it is maximal at 
$T\sim 2\times 10^4$~K (the ratio of the rates in photoequilibrium to that in the CIE reaches a factor of 5). 
In general, further increase in the number density leads to the domimation of collisions in wider temperature
range.
}
It is interesting that for $n=10^{-2}$~cm$^{-3}$ the contributions from carbon and nitrogen are higher 
than these for $n=10^{-4}$~cm$^{-3}$, e.g. for carbon the increasing factor reaches about a factor of 5 
around $T\simeq 7\times 10^5$~K, where it provides the maximum contribution.

\subsubsection{Importance of nonequilibrium effects}

\begin{figure*}
\includegraphics[width=18cm]{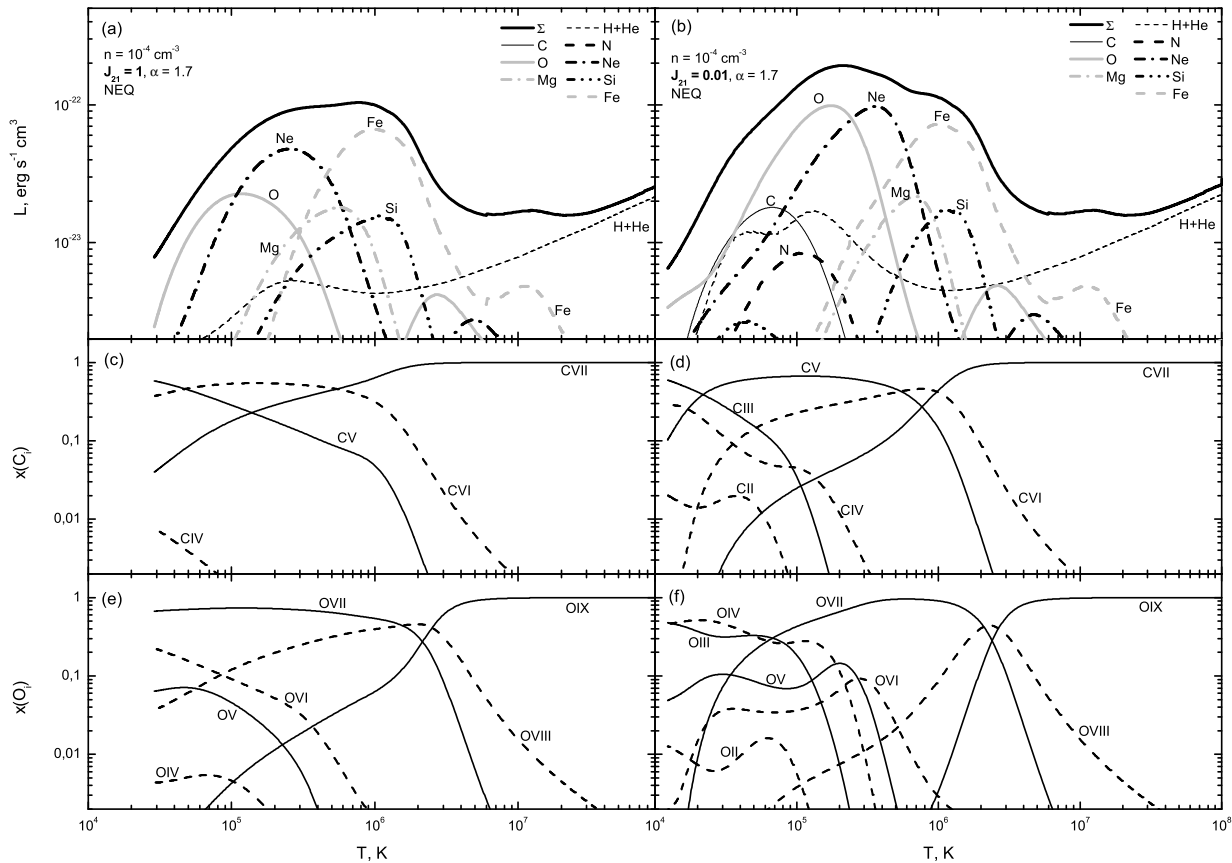}
\caption{
The cooling rates, carbon and oxygen ionization states in photo-nonequilibrium for solar metallicity 
and $n=10^{-4}$~cm$^{-3}$ exposed to the power-law ionizing radiation with $J_{21}=1$ (left panels)
and $J_{21}=0.01$ (right panels), the spectral index $\alpha$ equals 1.7. 
The thick solid black line shows the total cooling rate, the thin dash line shows the contribution from 
hydrogen and helium, the other lines (see the legend on panels) show the contributions from each
chemical element.
}
\label{figcoolantspl}
\end{figure*}

In order to determine a parameter space, where nonequilibrium effects play a significant 
role we calculate two sets of models: one is for power-law spectra, $\alpha=1.7$ and 5, 
the other is for the HM01 extragalactic background. 

Figure~\ref{figrelpl} presents the ratio of the nonequilibrium to equilibrium cooling rates for 
gas exposed to a power-law spectrum with $\alpha=1.7$ (panels (a-c)) and $\alpha=5$ (panel (d)). 
{ 
To compare easily the ratio of the cooling rates of collisional and photoionized plasma the ratio 
of collisional gas (no photoionization) is shown by the letter "c" on the y-axis (the area is 
separated by dash line) on each panel of Figure~\ref{figrelpl}. 
}
In the panels (a) and (b) the ratio is shown for $Z=10^{-3} Z_\odot$ and the solar metallicities.
As one expects the photoequilibrium and time-dependent cooling rates are almost equal for the lower 
metallicity. Only for both $J_{21}\simlt 10^{-3}$ and $T\simlt 10^5$~K the rates demonstrate a 
difference, which originates from the efficient H and He recombination in the photoequlibrium for 
such a low ionizing flux. For solar metallicity { (panel b)} the deviation between the rates can be found in 
wider range of both flux and temperature. The time-dependent rates tend to the photoequlibrium 
at $J_{21}\simgt 1$ (e.g. the ratio is close to 1) that corresponds to the ionization parameter log~$U\simgt -0.4$ 
(see the extra $y$-axis). The increase of density, $n=10^{-2}$~cm$^{-3}$, leads to further expansion of the range 
{ (panel c)}, where the photoequilibrium and photo-time-dependent rates show a significant difference. Under such 
conditions the photo-nonequilibrium rate tends to that in the collisinally cooling gas (the region 
below the dashed line). So the maximum difference between photoequilibrim and photo-nonequilibrium 
rates corresponds to that between the CIE and collisinally time-dependent cooling rates (note that
a significant difference between the collisional equilibrium and nonequlibrium rates at $T\simlt 2\times 10^4$~K
is determined by the efficient hydrogen recombination in the CIE). Similar 
difference can be seen in the panel (d), where the ratio is presented for the steeper spectrum,
$\alpha=5$. Note that in the panels 
{  
(b) and (d) the ratios are presented for the different power-law index $\alpha$, so the difference between 
the ratios at a given U for the power-law spectra with $\alpha=1.7$ and 5 is due to the different photon
distribution (in general, the dependence of cooling rates on ionization parameter can be considered if the 
same photon distribution is used). This difference is clearly seen for log~$U = -1.4$ and can be explained 
by stronger overionization of high ionic states by hard (far UV and X-ray) photons and longer recombination 
lag for the spectrum with $\alpha=1.7$ in comparison with that for the steepper spectrum with $\alpha=5$.
The number of such hard photons is larger for $\alpha=1.7$ than for $\alpha=5$, although the number of 
the hydrogen ionizing photons is the same for both spectra.
}

Panels (a-b) in Figure~\ref{figcoolantspl} present the individual contributions from each chemical 
element into the total cooling rate for gas with solar metallicity exposed to the power-law ionizing 
radiation with $J_{21}=1$ (a) and $J_{21}=0.01$ (b), the spectral index $\alpha$ equals 1.7. At 
$T\simgt 2\times10^6$~K the total cooling rates are very close to each other as well as almost equal 
to the photoequilibrium cooling rates (see Figure~\ref{figrelpl}b). In this temperature range the H, 
He and Fe ions provide the main contribution to the rate. {  At such a high temperature contribution 
of H and He is due to thermal bremsstrahlung.} Below this temperature the cooling by the Fe, 
Ne, O and C ions dominate. The contributions from Fe and Ne are significant in the ranges 
$T \simlt (0.6-2)\times 10^6$ and $T \simlt (1-6)\times 10^5$, correspondingly, for both fluxes considered 
here. If the flux (the ionization parameter $U$, in general) increases, then the Ne cooling bump strongly 
diminishes. This trend can be clearly seen in the panels (a) and (b). Further increase of the ionizing 
flux leads to the Fe bump diminishing. In Figure~\ref{figcoolantspl} similar situation can be seen for 
the O and espesially for C cooling bumps at $T\sim 10^5$~K. Actually, for $J_{21}=1$ the contribution 
from oxygen dominates only at $T$ below $10^5$~K, but for $J_{21}=0.01$ 
oxygen becomes a major coolant at $T\simlt 2\times 10^5$. Carbon demonstrates more remarkable increase 
of the cooling rate: for $J_{21}=1$ its contribution is negligible, but for $J_{21}=0.01$ it gives a 
significant part of the cooling rate and even becomes one of the dominant coolant at $T \sim (2-4)\times 10^4$~K. 
We note that for $J_{21}=1$ the difference of cooling rate between photoequilibrium and nonequilibrium 
is low as $\sim 10$\%, so the ionic composition is expected to be close to the photoequilibrium: the Ne 
and Fe ions are slightly overionized. The opposite situation is for $J_{21}=0.01$, the difference of 
the cooling rates reaches a factor of two at $T\simeq 2\times 10^5$~K. Under such a flux the ions of the main 
coolants (C, O, Ne and Fe) are strongly overionized in the range $T \simlt 10^6$~K, where the difference of 
the cooling rates becomes significant (see Figure~\ref{figrelpl}b). The Mg and Si give a minor contribution 
in the whole temperature range for two fluxes considered. 

Figure~\ref{figcoolantspl}(c-f) present the carbon and oxygen ionization states in a gas
exposed to the power-law ionizing radiation with $J_{21}=1$ (left panels) and $J_{21}=0.01$ (right panels)
with the spectral index $\alpha = 1.7$. For $J_{21}=1$ the increase of the cooling by oxygen at 
$T \simlt 5\times 10^5$~K is certainly determined by OVI and OV ions (see panel (e)). The contribution 
from oxygen at $T \simlt 5\times 10^4$~K is probably to go down due to the existence of the high ionic 
states only, OV-OVII, whose cooling rate is inefficient in this temperature range. 
Carbon is strongly ionized and the CV-CVII ions do not provide any significant energy losses.
For the lower flux, $J_{21}=0.01$, the photoionization rate is smaller and the recombination becomes more
efficient. The contribution from carbon at $T \simlt 2\times 10^4$~K is determined by the rapid increase
of the CIII and CIV ion abundances. The dominant contribution from oxygen in the wide temperature range
can be explained by the OV-OVI ions at $T \simgt 10^5$~K and the OIII-OIV ions at $T \simlt 10^5$~K.
Note that for a photoionized gas the decrease of metallicity leads to increase of the H and He contribution 
to the cooling rate. The cooling of low metallicity gas, $Z\sim 10^{-3}Z_\odot$, controlled by H and He 
ions only, the contribution from other elements is negligible.

\begin{figure}
\includegraphics[width=9cm]{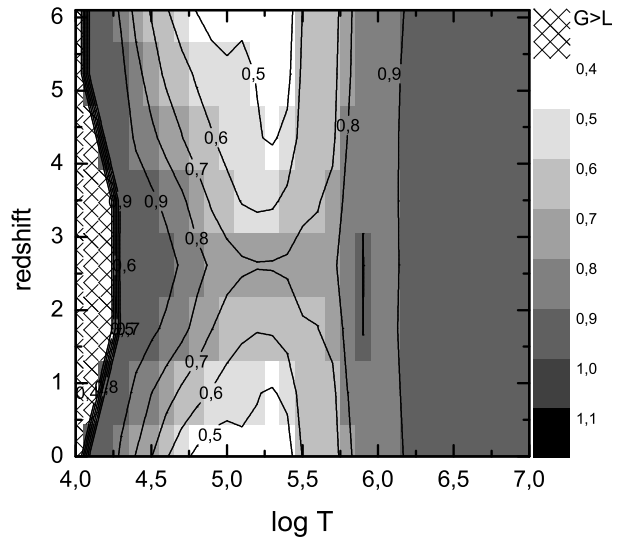}
\caption{
The ratio of nonequilibrium to equilibrium cooling rates for gas with density
$n=10^{-4}$~cm$^{-3}$ and solar metallicity exposed to the Haardt \& Madau (2001) 
background spectrum at redshift $z$. 
The shaded area corresponds to the conditions under which the photoheating exceeds the 
cooling in the nonequilibrium case.
}
\label{figrelhm}
\end{figure}

Figure~\ref{figrelhm} demonstrates the ratio for the HM01 background spectrum at redshift $z$. 
For the extragalactic HM01 background the picture is similar to that for a power-law spectrum: 
an increase of the metallicity leads to an increase of the magnitude of deviations from the 
equilibrium. For low metallicity, $Z\sim 10^{-3}~Z_\odot$, the ratio remains close to unity 
in the wide range of temperature. Only for $T\sim 10^5$~K it deviates from unity about 20-30\%
that is determined by the overionization of helium (see Figure~\ref{figcool}f).  
Figure~\ref{figrelhm} shows the ratio for solar metallicity. As one expects the nonequilibrium 
effects are important for $T\simlt 10^6$~K  in the whole redshift range, $z = 0-6$. The maximum
deviation reaches a factor of $\sim 2$ at $T\sim 10^5$~K for $z\simlt 1$ and $z\simgt 5$.
As the extragalactic background is maximal around redshifts $z = 2-3$ a saddle point on  
the surface of the cooling rate ratio is observed: the deviation from the equlibrium is minimum 
in this redshift range. 
As we mentioned above the difference from photoequilibrium is determined by smaller contribution
to the cooling from C, N, O and Ne ions, because they are overionized. The ionizing flux in
$z=2-3$ is the highest, so the ionic composition is moved to photoequlibrium (see the 
previous section). For both $z\sim 0$ and $z\sim 6$ the ionizing flux decreases and the 
difference from photoequilibrium becomes more clear.

Let us summarize where the photoequilibrium calculation can be used and how the parameters 
(metallicity, temperature, density and UV intensity) should be changed so that the nonequilibrium
effects become important. The calculations presented above show that the photoequilibrium
is reached in high-temperature gas, $T\simgt 3\times 10^6$~K, for any metallicity or ionization
parameter. 
{ 
Note that the photoequilibrium equals to the CIE for such a high temperature.
}
Below $T\sim 3\times 10^6$~K the deviation from the photoequilibrium grows.
For a metallicity as low as $10^{-3}Z_\odot$ such a deviation is minimum. The increase of 
metallicity leads to the widening of the parameter space, where nonequilibrium effects are 
significant (Figures~\ref{figrelpl} and \ref{figrelhm}), that is clearly seen for solar metallicity. 
This can be easily explained by the overionization of metal ions and the increase of the contribution 
from metals to the cooling rate for higher metallicity. For collisionally ionized
gas the deviation from the equlibrium at $T\simlt 10^6$~K is significant for any metallicity.
In a gas exposed to ionizing radiation the photoequilibrium is reached faster than higher
ionization parameter. At low ionization parameter (low flux intensity/high density), the 
differences between equilibrium and nonequilibrium indeed become large, but these 
are just the differences in the collisional case.
Thus, in Table~\ref{table2} for $T<10^6$~K we provide the parameter range under which 
the cooling rates in the equilibrium (EQ) and nonequilibrium (NEQ) are close, and a change
of parameters, which violates this equality. Here we should also pay attention on that the ionization parameter 
$U$ does not fully describe the spectrum and in general the transition from photoequilibrium to 
nonequlibrium depends on the ionizing spectrum shape.

As a conclusion we note that the deviation of the cooling rates between photoequilibrium and 
nonequilibrium is determined by recombiantion lag in the latter case. The maximum difference 
is for collisionally controlled gas. Such condition in gas is reached for ionization parameter as low 
as log~$U\simlt -5$ (see Figures~\ref{figrelpl}--\ref{figrelhm}). The difference decreases 
with the ionization parameter growth. For sufficietly high ionization parameter the photoequlibrium
can be reached. However, the transition region from photoequilibrium to nonequilibrium 
is quite wide. For ionization parameter as low as log~$U\sim -5$ the nonequilibrium collisional 
cooling rates and ionic composition are a better choice. The photoequilibrium is reached for 
ionization parameter as high as log~$U\simgt -2$ for $Z\simlt 10^{-2}Z_\odot$ and log~$U\simgt 0$ 
for $Z\sim Z_\odot$ (see Table~\ref{table2}). For supersolar metallicity the photoequilibrium 
is expected to reach for higher ionization parameter. So where the physical conditions are favour to 
collisional ionization (the collisions dominate or its contrubution to the total ionization rate 
is significant), the nonequilibrium (photo)ionization calculations should be conducted. 

\begin{table}
\caption{The transiton from photo-EQ to NEQ at $T<10^6$~K.}
\center
\begin{tabular}{ll}
\hline
\hline
{  the conditions, where} & {  a variation of parameters} \\
{  $\Lambda^{EQ}\simeq \Lambda^{NEQ}$ is expected}  & {  leads to $\Lambda^{EQ}\neq \Lambda^{NEQ}$ }     \\   
\hline
\ high ionization parameter $U$                         & \ {\it either} increase of $Z$ \\
\ \ (high UV flux/low density),                       & \ {\it or} decrease of $U$ parameter  \\
\ \  $U\simgt 10^{-2}$ for $Z\simlt 10^{-2}Z_\odot$ , & \ \  (decrease of flux/increase  \\
\ \  $U\simgt 1$ for $Z\sim Z_\odot$                  & \ \   of density) \\
\hline
\hline
\end{tabular}%
\label{table2}
\end{table}

\subsection{Ionization states of metals and their ratios}

\subsubsection{Collisions or photoionization: CIII and CIV ions}

\begin{figure}
\includegraphics[width=9.3cm]{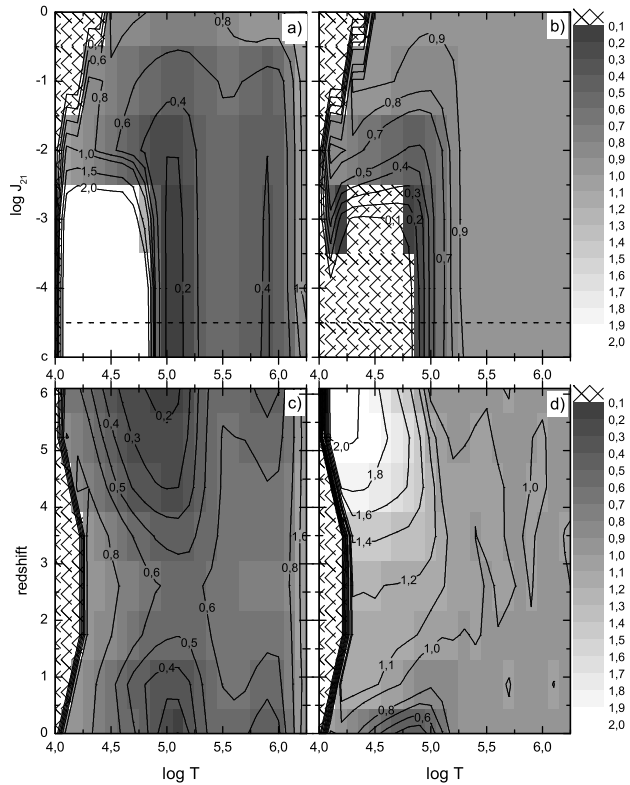}
\caption{
The ratio of the CIV ion fraction in the nonequilibrium case to that in the equilibrium (left panels),
and the ratio, $R = x({\rm CIII})/x({\rm CIV})$, in the nonequilibrium to that in the 
equilibrium (right panels) for gas with density $n=10^{-4}$~cm$^{-3}$ and solar metallicity 
exposed to the power-law spectrum with $\alpha=1.7$ (upper row) and the \citet{hm01} 
background spectrum at redshift $z$ (lower row). 
In the panels (a) and (c) The letter "c" on the y-axis (or the area is separated by dash line) 
corresponds to the ratio of the cooling rates of collisional plasma (no photoionization). 
The shaded area corresponds to the conditions under which the the ratio does not exist 
(photoheating exceeds the cooling in the nonequilibrium case) or the ratio is out of 
the ranges presented in the gray scale.
}
\label{figrelri}
\end{figure}

In the previous sections we have considered the time-dependent cooling functions of a photoionized gas,
their deviations from the photoequilibrium rates and the contribution to the cooling rate from chemical 
elements. We have found that the transition region from photoequilibrium to nonequilibrium is quite wide.
Obviously, the reason of such a transition is the recombination lags in the nonequilirium 
model. Consequently, a significant difference of the ionic ratios is expected between the photoequilibrium and the
nonequilibrium. Moreover, which of two processes, collisions or photoionization, is dominant can be also
identified using ionic ratios. For gas with a given density and metallicity exposed to an ionizing radiation 
the transition between collisionally controlled or photoionization regimes depends on the photoionization 
rate, which is determined by spectrum shape and flux magnitude. 

Figure~\ref{figrelri} shows the ratio of the CIV ion fraction in nonequilibrium to that in photoequilibrium 
(left panels) for gas with density $n=10^{-4}$~cm$^{-3}$, solar metallicity exposed to the power-law 
spectrum with $\alpha=1.7$ (upper) and the HM01 background spectrum at redshift $z$ (lower). 
As one expects the CIV fraction in the photoequilibrium strongly differs from that in the time-dependent 
case: the ratio reaches several times. The significant deviation from photoequilibrium can be also found for 
the HM01 background. In the panel (a) one can see that the decrease of the UV flux amplitude leads to that 
the CIV fraction ratio tends to that in collisionally controlled gas. 
We should note once again that due to the recombination lag the carbon ionic states in the nonequilibrium are
always overionized in comparison with these in the equilibrium (see the panels (a) and (c)): where the ratio 
is below 1, the CV state dominates in the nonequilibrium, e.g. the CV is overionized, and in the opposite case, 
where the ratio is greater than 1, the CIV is overionized, whereas in the photoequilibrium under the same conditions 
the CIII fraction increases fastly. This is clearly seen in the right panels of Figure~\ref{figrelri}, where 
the ratio of the values, $R = x({\rm CIII})/x({\rm CIV})$, in the nonequilibrium to that in the equilibrium 
is presented. Indeed, in the panel (b) the shaded region in the bottom left conner corresponds to the ratio less 
than 0.1, i.e. the denominator of the ratio, the $R$ value in the equlibrium, is high.

In the panel (d) the strong deviations of the $R$ value from the photoequilibrium can be found for the HM01 spectrum.
Because the CIII/CIV ratio is used as one of the important observable indicator to discriminate between the HeII 
reionizaton models \citep{schaye03,agafonova,madau09}, we should pay attention on the significant difference at
redshifts $z=2-3$, when the HeII reionization is expected to occur. The nonequilibrium analysis of the CIII/CIV 
ratio in the absorption spectra obtained by \citep{agafonova} can be found in \citep{v10}. Certainly, for lower
metallicity smaller deviations are expected. However, one can conclude that the nonequilibrium effects are 
significant and it should be taken into account in the simulations of the IGM state.

\subsubsection{An example of the ratio of column densities}

\begin{figure}
\includegraphics[width=8.5cm]{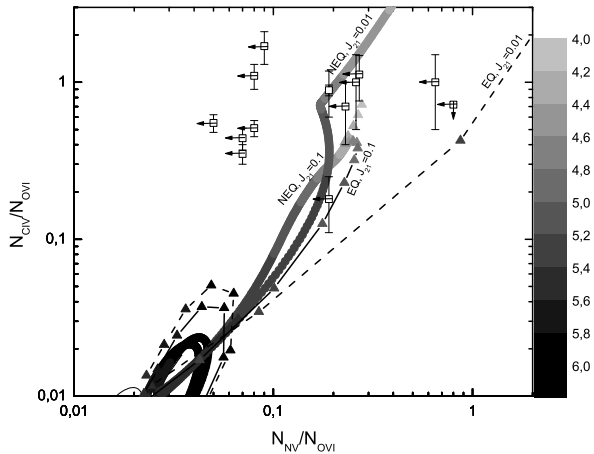}
\caption{
The dependence $N_{\rm CIV}/N_{\rm OVI}$ versus $N_{\rm NV}/N_{\rm OVI}$
in the photo-nonequilibrium model for cooling gas with $n=10^{-4}$~cm$^{-3}$ 
and $Z = Z_\odot$ exposed to the ionizing radiation with $\alpha=1.7$ and 
two fluxes $J_{21}=0.01$ and 0.1 (the corresponding labels are shown).
The same dependence in the photoequilibrium are by triangles connected by solid line
for $J_{21}=0.1$ and by dashed line for $J_{21}=0.01$. Gas temperature is indicated 
by gray scale along the trajectories: from hot (black) to cold (light gray) gas.
The data points show the ionic ratios observed in metal absorbers \citep[see Table 8][]{hvc1}.
}
\label{figratio1}
\end{figure}

The ratio of column densities of ion $i$ of element $m$ and ion $j$ of element $n$ is \citep{gs07}
\be
{N_i^m \over N_j^n} = {A_m\over A_n}{x_i(T) \over x_j(T)},
\ee
where $A_m, A_n$ are the abundances of elements $m$ and $n$ (relative to H), 
$x_i, x_j$ are the fractions of corresponding ionization states. 
The absence of the hydrogen density dependence provides more general tool for diagnostics of the IGM. 
Following \citet{gs07} we consider the ratios $N_{\rm CIV}/N_{\rm OVI}$ and $N_{\rm NV}/N_{\rm OVI}$. 

Figure~\ref{figratio1} shows these ratios for power-law spectrum with $\alpha=1.7$ and
two fluxes $J_{21}=0.1$ and 0.01 in photoequilibrium and photo-nonequilibrium. 
The "cooling tracks" are the gray diagrams, where the gray gradation corresponds to gas 
temperature.  
As can be expected, the strong difference between the photoequilibrium and time-dependent tracks
is seen for $J_{21}=0.01$, whereas the tracks for $J_{21}=0.1$ become closer. 
Indeed, the ionic ratios in a gas exposed to strong ionizing radiation significantly differ 
from the ratios in collisionally controlled gas (see e.g. Figure~\ref{figrelri}). The decrease 
of ionizing flux increases the role of collisions and leads to the transition between photoionized 
and collisional plasma. However, as one can see in Figures~\ref{figrelpl} and \ref{figrelhm}-\ref{figrelri} 
there is no sharp transition between photoionized and collisional plasma. So ionic ratios observed 
in an object can be satisfied by both photoionized and collisional time-dependent conditions,
{ 
e.g. for $J_{21}=0.001$ the ratio, $R=x({\rm CIII})/x({\rm CIV})$, demonstrates a negligible difference
from the collisional time-dependent ratio (see Figure~\ref{figrelri}b). But the difference becomes very clear for $J_{21}=0.01$ at $T\simlt 10^5$~K.
}

Figure~\ref{figratio1} shows the observational data for the high velocity clouds in the Galactic halo 
\citep[see Table 8 in][]{hvc1}. \citet{gs07} showed that the column density ratios in the high velocity clouds 
observed by \citet{hvc1} can be fitted using the nonequilibrium collisional model. Here we confirm that the nonequilibrium photoionization models can also explain the observed ratios, for a gas with $n\sim 10^{-4}$~cm$^{-3}$
exposed to $J_{21}=0.01-0.1$ and $\alpha=1.7$ in particular. For example,
the ionization states of carbon and oxygen for $J_{21}=0.01$ can be found in Figure~\ref{figcoolantspl}.
Under such a flux the CIV, NV and OVI fractions are close to these in the collisional time-dependent model, 
but strongly overionized in comparison with the photoequilium ones. More detailed study is out of scope of this paper.

\section{Conclusions}

\noindent

In this paper we have presented the nonequilibrium cooling rates and 
ionization states of gas enriched with heavy elements and photoionized by external 
ultraviolet/X-ray radiation. We have studied the dependence of cooling rates on the gas density 
and metallicity, and also by the radiation field \citep[power-law and the extragalactic background spectra
obtained by][]{hm01}. 
The calculations are done for a wide range of physical parameters: $10^4 - 10^8$~K in the gas temperature, 
$Z = 10^{-3} - 1~Z_\odot$ in the metallicity.
For the power-law spectrum, the spectral index $\alpha$ is taken equal to 1, 1.7 and 5, and 
for the extragalactic background spectra the redshift varies from 0 to 6. 
In all calculations, we assume that the gas is optically thin.
We have found the following.
\begin{itemize}
 \item The time-dependent cooling rates depend significantly on the gas metallicity and density
 as well as on the flux and shape of the ionizing radiation.
 \item The time-dependent cooling rates and ionic composition of gas 
 differ strongly from those in the photoequilibrium due to the overionization
 of ionic states in the time-dependent case. The time-dependent 
 cooling efficiency is generally lower than that in the photoequilibrium, and
 the nonequilibrium effects can be clearly seen at $T\simlt 10^6$~K for the solar
 metallicity.
 \item The difference between the time-dependent and equilibrium rates can be significant, 
 its magnitude is maximal at ionization parameter as low as log$U\simlt -5$, and similar to 
 the difference between the equlibrium and the nonequilibrium cooling rates in the collisionally 
 controlled gas (see Figures~\ref{figrelpl}--\ref{figrelhm}).
 \item Contributions by individual chemical elements to the total cooling rate can be 
 significantly different for time-dependent and equilibrium plasma at $T\simlt 10^6$~K.
 \item The mismatch of both ionic states and their ratios in the photoequilibrium and photo-nonequilibrium
 can reach a factor of several.
\end{itemize}
In this paper we have considered the external radiation and gas parameters, under which 
the cooling rates and ionic composition in the photoequilibrium and photo-nonequilibrium 
are expected to be close to each other, and a change of the parameters which violates this equality. 
The parameter space where the nonequilibrium effects are significant is quite wide. Because 
the difference between the photoequlibrium and the nonequilibrium is maximal for the ionization
parameter log~$U \simlt -5$, then the nonequilibrium collisional cooling rates and ionization states 
are a better choice for such a low ionization parameter. We have found that the difference decreases with the 
ionization parameter growth, and the photoequilibrium is reached for ionization parameter 
as high as log~$U\simgt -2$ for $Z\simlt 10^{-2}Z_\odot$ and log~$U\simgt 0$ for $Z\sim Z_\odot$ 
(see Table~\ref{table2}). For supersolar metallicity the photoequilibrium is expected to reach 
for higher ionization parameter. 
We have concluded that the nonequilibrium photoionization calculations should be conducted,
when the contribution from collisions to the total ionization rate is significant. 
The main reason of the deviation in the time-dependent case is the recombination lag, so 
the ionic states are expected to be overionized, especially this is notable for metals.
Therefore we should pay attention on the importance of use nonequilibrium cooling rates for 
near-solar (and above) metallicity of gas exposed to arbitrary ionizing radiation flux. 
Thus, the nonequilibrium cooling rates are the necessary component for study metals 
in the IGM, mixing metal processes, damped Lyman-$\alpha$ systems, galactic superwinds and so on. 
Also to simulate correctly the WHIGM ionic composition we need nonequilibrium processes and more 
exact models for ionizing background flux.

\section{Acknowledgements}

\noindent

The anonymous referee is acknowledged for valuable comments and criticism.
The author is grateful to: Yuri Shchekinov for help and many useful discussions; 
Enn Saar, Sviatoslav Dedikov, Eugene Kurbatov, Eugene Matvienko, Alexei Shaginyan and 
Eduard Vorobyov for discussions, their help and advices.
Gary Ferland and CLOUDY community are acknowledged for creating of the excellent tool
for study of the photoionized plasma -- CLOUDY code.
This work is supported by the RFBR (project codes 08-02-91321, 09-02-90726, 09-02-00933, 
10-02-90705 and 11-02-90701),
by the Federal Agency of Education (project code RNP 2.1.1/1937)
and by the Federal Agency of Science and Innovations (project 02.740.11.0247). 
The author acknowledges the Special Astrophysical Observatory RAS and the Institute of Astronomy
RAS for hospitality, where he was a visiting scientist.


\end{document}